\documentclass{article}

\usepackage[numbers]{natbib}
\usepackage[final]{neurips_2019_ml4ps}

\usepackage[utf8]{inputenc} 
\usepackage[T1]{fontenc}    
\usepackage{hyperref}       
\usepackage{url}            
\usepackage{booktabs}       
\usepackage{amsfonts}       
\usepackage{amssymb}
\usepackage{nicefrac}       
\usepackage{microtype}      
\usepackage[pdftex]{graphicx}

\pdfoutput=1

\title{
Using U-Nets to Create High-Fidelity Virtual Observations of the Solar Corona
}

\author{Valentina Salvatelli  \\
  NASA Frontier Development Lab \& IQVIA\\
  \And
  Souvik Bose\\
  Rosseland Center for Solar Physics, \\ University of Oslo
  \And
  Brad Neuberg\\
  NASA Frontier Development Lab\\
  \And
  Luiz F. G. dos Santos\\
  IACS - The Catholic University of America \\ 
  NASA - Goddard Space Flight Center\\
  \AND
  Mark Cheung\\
  Lockheed Martin Solar and Astrophysics Laboratory, \& \\
  Stanford University \\
  \And
  Miho Janvier\\
  Institut d'Astrophysique Spatiale, Université Paris-Sud\\
  \And
  Atilim Gunes Baydin\\
  University of Oxford\\
  \And
  Yarin Gal \\
  OATML, University of Oxford \\
  \And
  Meng Jin\\
  Lockheed Martin Solar and Astrophysics Laboratory \& SETI\\
  }

\begin{document}

\maketitle

\begin{abstract}
  Understanding and monitoring the complex and dynamic processes of the Sun is important for a number of human activities on Earth and in space. For this reason, NASA’s Solar Dynamics Observatory (SDO) has been continuously monitoring the multi-layered Sun's atmosphere in high-resolution since its launch in 2010, generating terabytes of observational data every day. The synergy between machine learning and this enormous amount of data has the potential, still largely unexploited, to advance our understanding of the Sun and extend the capabilities of heliophysics missions. In the present work, we show that deep learning applied to SDO data can be successfully used to create a high-fidelity “virtual telescope” that generates synthetic observations of the solar corona by image translation. Towards this end we developed a deep neural network, structured as an encoder-decoder with skip connections (U-Net), that reconstructs the Sun's image of one instrument channel given temporally aligned images in three other channels. 
  The approach we present has the potential to reduce the telemetry needs of SDO, enhance the capabilities of missions that have less observing channels, and transform the concept development of future missions.
\end{abstract}

\section{Introduction}
\label{intro}
Solar activity deeply affects our life on Earth and activities in space. In order to monitor and understand the solar system dynamics, NASA operates the Heliophysics System Observatory (HSO), a fleet of satellites that constantly observe the Sun, its extended atmosphere and the space environments around Earth. One of its missions, the Solar Dynamics Observatory~\cite{SDO_primary},  has been monitoring the million Kelvin solar corona 24x7 in 9 different extreme UV (EUV) wavelength bands since 2010 and it alone produces terabytes of observational data every day. This enormous amount of data presents analytics challenges but also opportunities for scientific discovery and for enhancing the capabilities of the HSO. For example, future HSO missions will be required in order to understand the fundamental processes of our star, especially deep space missions that will observe the Sun from alternate vantage points. However, due to the low telemetry rates and the high cost of using the Deep Space Network, deep space missions cannot transmit data at the full spatial, spectral and temporal sampling rate of the on-board sensing system. 

One approach to tackle this problem is to produce proxy science data products using alternate data sources. In this paper, we address the specific question: could EUV images of the solar corona at a certain wavelength be synthesized using images acquired at other wavelengths? 

Our hypothesis is that, from the petabyte-scale archive of images accumulated by SDO, a deep learning model can learn  the correlations that exist between the spatial features of the solar corona as measured by different wavelengths, and exploit these correlations for image translation.

\section{Methodology}
\label{method}

\subsection{Data Description}
The virtual telescope presented in this work is based on data from SDO's Atmospheric Imaging Assembly (AIA)~\cite{AIA}. The AIA instrument takes full-disk, 4096 by 4096 pixel, imaging observations of the solar photosphere, chromosphere and corona in 2 UV channels and in 7 extreme UV (EUV) channels. The original SDO dataset was processed in \cite{SDOML} into a machine-learning ready dataset of \textasciitilde{}6.6 TB (hereafer SDOML) that we leveraged for the current work. 

The SDOML dataset is a subset of the original SDO data ranging from 2010 to 2018. Images are spatially co-registered, have identical angular resolutions, are corrected for the instrumental degradation over time and have exposure corrections applied. All the instruments are temporally aligned. AIA images in the SDOML dataset are available at a sampling rate of 6 min. The 512x512 pixel full-disk images have a pixel size of $4.8$ arcsec. The images are saved in single-precision floating point to preserve the high dynamic range ($\gtrsim 14$ bits per channel per pixel).

\subsection{Approach} \label{sec.app}
\begin{figure}[!htb]
  \centering
  \includegraphics[width=0.85\linewidth]{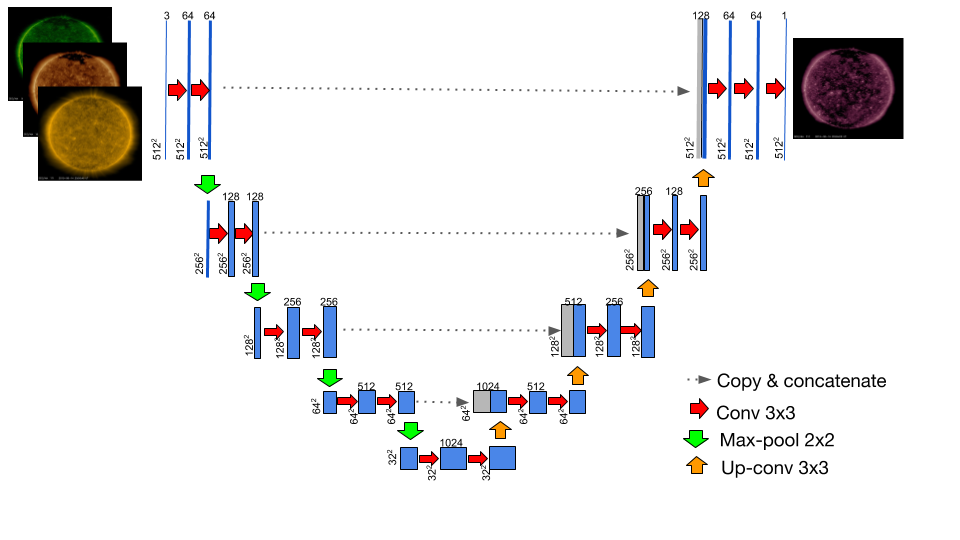}
  \caption{U-Net based architecture used to synthesize solar EUV images. Each box corresponds to a multi-channel feature map. Grey boxes are copied maps. The number of channels is shown on top of the box. Resolution in pixels is indicated on the left of the box. Arrows represent operations.}
  \label{fig.u-net}
\end{figure}

Our approach of synthesizing solar EUV images is to create a "virtual telescope" to perform image translation. Specifically, the virtual telescope we designed generates AIA EUV channel images, at a single time step, from measurements of other AIA channels at the same time step. Unlike many computer vision problems, it is not sufficient for scientific purposes to generate images that visually resemble the ground truth. This challenge requires the synthesis of high-fidelity images over a large dynamic range.  For the development of this work we focused on 4 channels (94, 171, 193, 211 \AA) that are sensitive to coronal plasmas at different temperatures~\citep{Cheung:2015}. We developed and trained a deep neural network (DNN) that takes input images from 3 channels and predicts the image of the 4-th channel, particularly we focused on the translation (94, 171, 193 \AA) to 211 \AA. The encoder-decoder approach is achieved by adopting a U-Net architecture \cite{Unet15}. The reason to choose a U-Net is its ability to exploit feature correlations at larger scales while preserving feature details at small scale. 
We used Leaky ReLU~\cite{Maas13rectifiernonlinearities} as the activation function between layers and, in order to mitigate the instabilities due to high contrast regions, we used a smooth L1 loss function~\cite{huber:1964}. A schematic of the architecture is given in Fig. \ref{fig.u-net}.

The high dynamic range of EUV emission from solar plasma implies that the pixel intensities span several orders of magnitude; for this reason we normalized the intensity values by the mean of the channel, rather than using standardization as usually done in other deep learning applications. We also did not apply batch normalization, discovering experimentally that this leads to better results.

In order to avoid biases due to the solar cycle, we split the dataset for training and testing according to the month, rather than the year. Namely, data between January and July are used for training. Data between August and October are used for testing. November and December are kept as holdout.

For model evaluation we used a metric that measures both pixel-by-pixel and structural differences between the predicted images and the ground truth images. As detailed in Eq.~\ref{metric}, this metric is the average of the root mean-squared error (RMSE) and the structural similarity index (SSIM)~\cite{Wang04imagequality}. Lower values mean better performance in this metric.

\begin{equation}
 {\rm Err}(I_1, I_2) = \frac{{\rm RMSE}(I_1, I_2) + \left[1-|{\rm SSIM}(I_1, I_2)|\right]}{2}
\label{metric}
\end{equation}

\subsection{Baseline model}
In order to compare the DNN with a simpler machine learning algorithm, we propose a baseline that corresponds to taking a linear regression of the three input channels: 
\begin{equation}
I_{\rm pred} = \alpha I_{1} + \beta I_{2} + \gamma I_{3} + \delta
\end{equation}
where $I$ represents the pixel intensity of the channels (subscripts for the three input channels); $\alpha$, $\beta$, $\gamma$ are 3 different factors expressing the different weights; and $\delta$ is the bias in the linear combination of the channels. 

\begin{figure}
  \centering
  \includegraphics[width=0.42\linewidth]{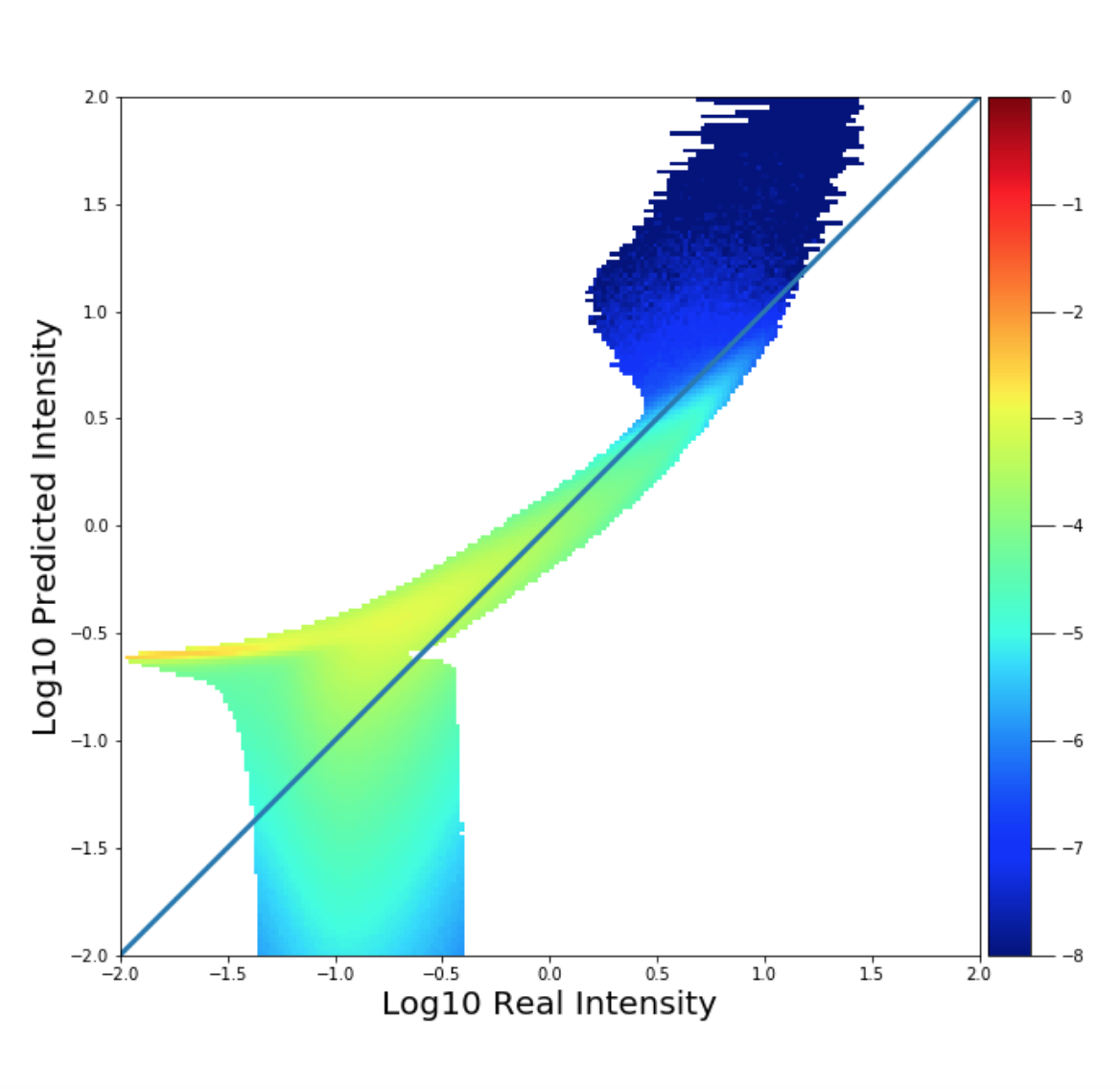}
  \includegraphics[width=0.43\linewidth]{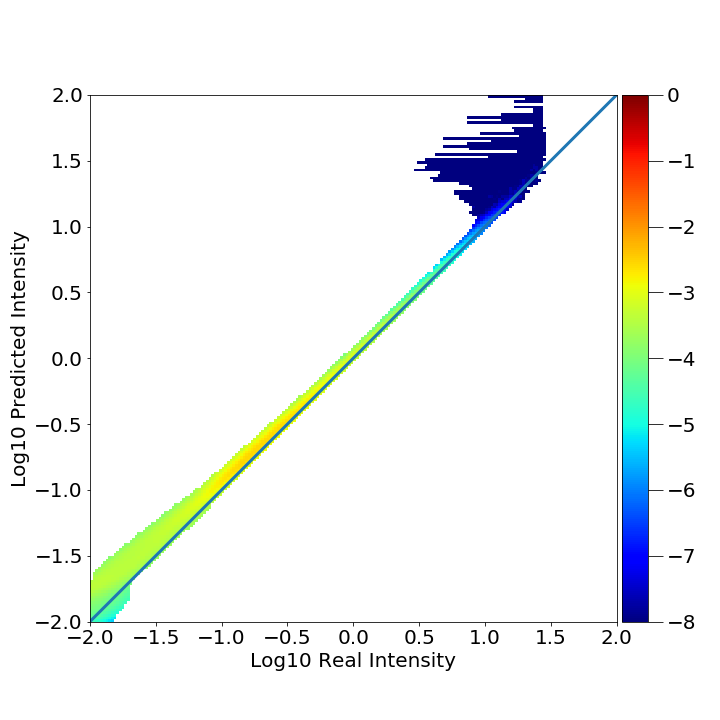}
  \caption{90\% confidence level for the conditional probability $P(I_{\rm GT} | I_{\rm pred})$ in the linear model (left) and in the U-Net model (right). The colorbar indicates the joint probability density function $\log_{10} {\rm JPDF} (I_{\rm GT}, I_{\rm pred})$. Dark blue are extremely rare data points.}
  \label{Baseline_DNN_cl}
\end{figure}

The results of the baseline model are shown on the left of Fig.~\ref{Baseline_DNN_cl}, which gives the joint probability density function (JPDF) of the ground truth and predicted intensities $P(I_{\rm GT}, I_{\rm pred})$. If the model were perfect, the JPDF would be confined to the diagonal line. The horizontal spread of ground truth ($I_{\rm GT}$) for a given predicted value ($I_{\rm pred}$) shows the 90\% confidence level for the conditional probability $P(I_{\rm GT} | I_{\rm pred})$. 

\section{Results}
\begin{figure}
  \centering
  \includegraphics[width=1.0\linewidth]{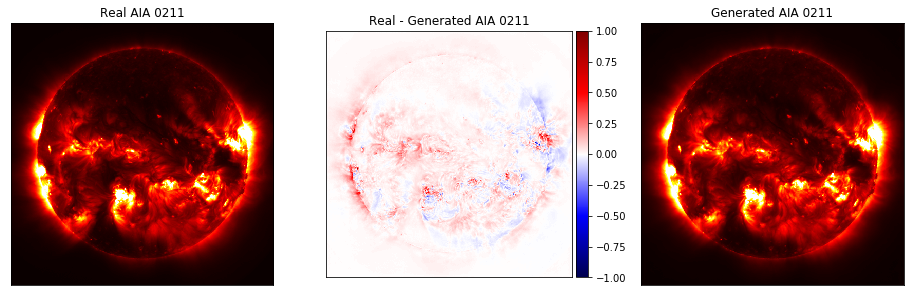}
  \caption{Left: original image in the 211 \AA{} channel. Right: synthetically generated image via U-Net. Middle: difference between real and generated image in units of 1000 data numbers/s/pixel. Brightest pixels on the limb are the most difficult to learn.}
  \label{fig.gt_real_diff}
\end{figure}

An example of resulting recovered image, when adopting the DNN architecture described in Sec.~\ref{sec.app}, is given in Fig.~\ref{fig.gt_real_diff}. In this figure, the left image corresponds to the original image in the 211 \AA{} channel, while the image on the right corresponds to the one generated by the DNN. Based on visual inspection, the synthetic image roughly reproduces the morphology of coronal loops in the ground truth image. This outperforms previous results in \cite{Park_2019}, where a conditional generative adversarial network (CGAN) had been trained to translate HMI magnetograms to AIA images. The result in unsurprising since the input AIA channels used in our U-Net have sensitivity to the plasma observed in the 211 \AA{} channel.

We also show in the central image of Fig.~\ref{fig.gt_real_diff} the differences between the real and generated images. Dark blue and bright red correspond to the regions where the differences are the largest, and can be seen to be located where the active regions (shown as the brightest regions in the original and generated images) are.

\begin{figure}
  \centering
  \includegraphics[width=0.9\linewidth]{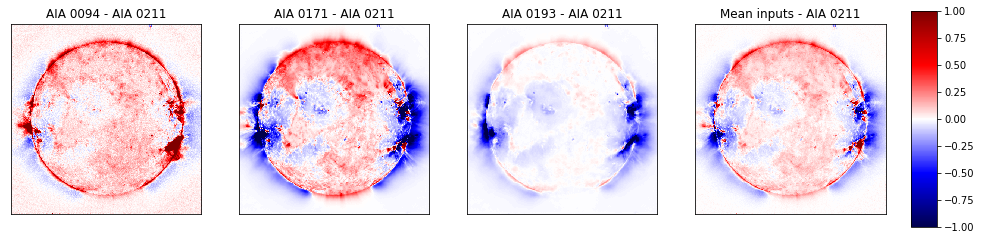}
  \caption{The first image represents the differences between the generated 211 \AA{} channel and the original 94 \AA{} channel, the second image the differences between the generated and the 171 \AA{} channel, the third image the generated and the 193 \AA{} channel, the fourth, the difference between the generated 211 \AA{} and the mean of the input channels. The DNN model is not dominated by one single channel.}
  \label{fig.input_output_diff}
\end{figure}

To understand whether these differences are predominantly from one channel, or are channel-independent, we provide in Fig.~\ref{fig.input_output_diff} the differences in the pixel brightness between the generated image and each of the three input channels images. The fact these differences all look different from the central image in Fig.~\ref{fig.gt_real_diff} confirms that the discrepancies are not dominated by one single channel. Also, the comparison of the generated image with the average of the three inputs, shown in the fourth image of Fig.~\ref{fig.input_output_diff}, indicates the neural network learns a non trivial combination of the input channels.

According to the metric defined in Eq.~\ref{metric} the baseline and the U-Net models have scores of 0.475 and 0.042, respectively. This demonstrates that the DNN improves by a factor of 10 over the linear model performance. As shown in Fig.~\ref{Baseline_DNN_cl} the improvement is homogeneous across 3 orders of magnitude in pixel brightness. Both the models cannot accurately reproduce extremely rare and bright pixels (dark blue area). However at high solar brightness, AIA suffers saturation of the detector, so the measured intensity values at the high end are not reliable. 

\begin{figure}
  \centering
  \includegraphics[width=1.0\linewidth]{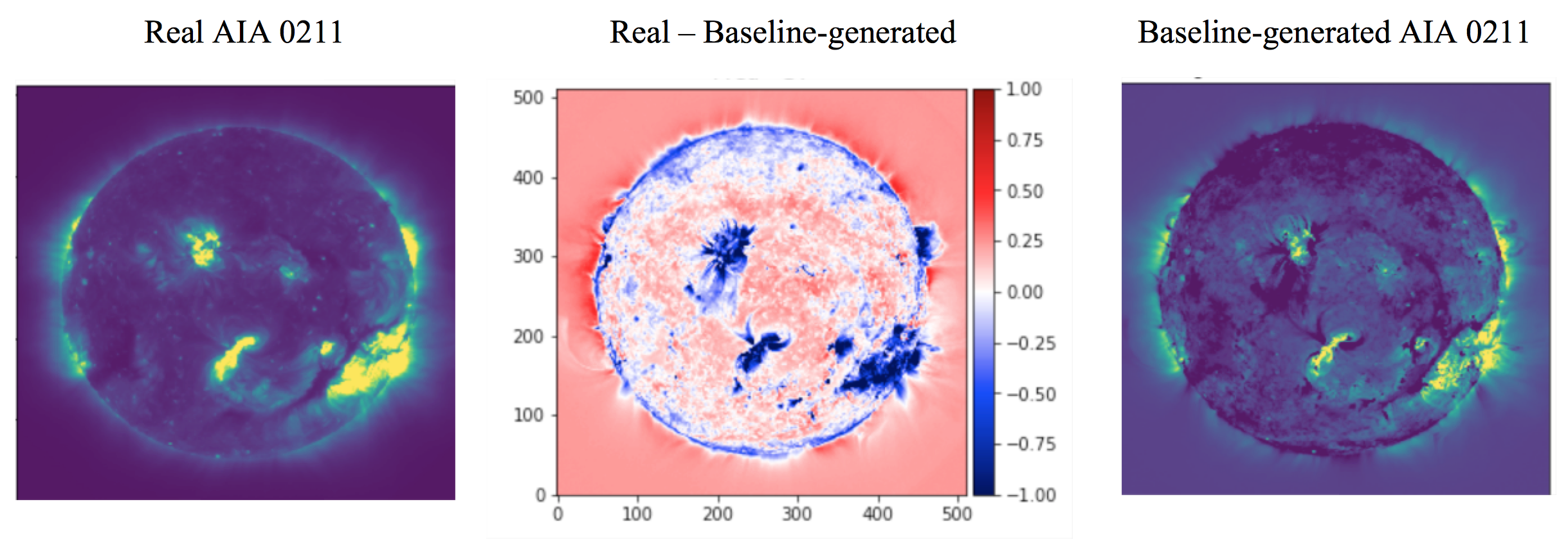}
  \caption{Left: original image. Right: image synthetically generated via linear model. Middle: difference between real and generated image. The units of the color scale are the same of the middle panel of Fig.~\ref{fig.gt_real_diff}. A more intense color means a bigger reconstruction error.}
  \label{image_baseline}
\end{figure}

 For comparison, an example of image reconstructed via the baseline linear model is presented in Fig.~\ref{image_baseline}, together with a color-map of the differences between the real and the predicted images.  In the difference map, the high color contrast and the dark blue areas in correspondence of active regions visually confirms, when compared with Fig.~\ref{fig.gt_real_diff}, the remarkable improvement introduced by the DNN over the baseline.

 We also explored the possible permutations of three input channels and one output channel, see Tab.~\ref{Tab-permutation}. We found that the same architecture produces similar or better reconstruction errors for 171 \AA{} and 193 \AA{}. The reconstruction error of 94 \AA{} is instead much higher. This is not surprising because its peak emission lies at a considerable higher temperature than the input channels (see Fig.~1 of \cite{Cheung:2015}). 

\begin{table}[h!]
\caption{Performance of the DNN on different permutations of input/output channels in the set (94, 171, 193, 211 \AA{}). In every column the input channels are all but the one produced as output.}
\label{Tab-permutation}
\centering
\begin{tabular}{|c||c|c|c|c|}
 \hline
 & \multicolumn{4}{|c|}{\textbf{Re-constructed AIA channel}} \\
 \hline
 \textbf{Metric} &  $211$ & $193$ & $171$ & $94$ \\
 \hline
 [ RMSE + (1-SSIM) ]/2    & 0.0424  &0.0369 &   0.0278 & 2.44 \\
 \hline
\end{tabular}
\end{table}

\section{Summary \& Future Work}
We hypothesized that multiple AIA channels could be used to synthesize a single missing AIA channel; we found this approach generates promising high-fidelity results, confirming that spatial correlations exist among the AIA channels and can be leveraged to reconstruct one channel from others. The suitability of such virtual telescope for scientific purposes will depend on the specific science use cases and will need further investigation to better understand the extent of the correlation and the limits of the model (i.e. range of pixel brightness that can be reproduced).

We plan to continue this work with the following refinements: testing the generality of our results by experimenting more broadly with channels combination; varying the architecture in order to find an ideal balance between performance and model complexity (simpler models would be advantageous for deployment); improving the predictions in active regions and flares, which can be achieved by including changes in the architecture and loss function; and finally, reproducing physical variables by using the synthetically generated images, as this will ultimately demonstrate the potential use of the virtual telescope for science.

\subsubsection*{Acknowledgments}
This project was conducted during the 2019 NASA Frontier Development Lab (FDL) program, a public/private partnership between NASA, SETI and industry partners including Lockheed Martin, IBM, Google Cloud, NVIDIA Corporation and Intel. The authors wish to thank in particular IBM and Google Cloud for generously providing computing resources. We gratefully thank all our mentors for guidance and useful discussion, as well as the SETI Institute for their hospitality during the program.

\bibliographystyle{plain}
\bibliography{virtual_telescope}

\end{document}